\documentclass[12pt]{article}
\usepackage{amsmath}
\usepackage{graphicx,psfrag,epsf}
\usepackage{enumerate}
\usepackage{natbib}
\usepackage{url} 
\usepackage{xcolor}
\usepackage{amsfonts}
\usepackage{amssymb}
\usepackage{booktabs}
\usepackage{rotating}
\usepackage{changepage}
\usepackage{amsthm}
\usepackage{bm}
\usepackage{hyperref}
\usepackage{quiver}
\usepackage{adjustbox}

\newcommand{\blind}{0}
\newtheorem{prop}{Proposition}
\newtheorem{coro}{Corollary}
\newtheorem{lemma}{Lemma}
\newtheorem{hyp}{A}
\newtheorem{theorem}{Theorem}
\newtheorem{defi}{Definition}
\begin{document}

\def\spacingset#1{\renewcommand{\baselinestretch}%
{#1}\small\normalsize} \spacingset{1}


\if0\blind
{
  \title{\bf Copula Central Asymmetry of Equity Portfolios}
  \author{Lorenzo Frattarolo\thanks{
     lorenzo.frattarolo@univr.it}\hspace{.2cm}\\
    Department of Economics, University of Verona, Italy. Email: 
    }
  \maketitle
} \fi

\if1\blind
{
  \bigskip
  \bigskip
  \bigskip
  \begin{center}
    {\LARGE\bf Copula Central Asymmetry of Equity Portfolios}
\end{center}
  \medskip
} \fi

\bigskip
\begin{abstract}
Financial crises are usually associated with increased cross-sectional dependence between asset returns, causing asymmetry between the lower and upper tail of return distribution. The detection of asymmetric dependence is now understood to be essential for market supervision,  risk management, and portfolio allocation.
  I propose a non-parametric test procedure for the hypothesis of copula central symmetry based on the Cram\'er-von Mises distance of the empirical copula and its survival counterpart, deriving the asymptotic properties of the test under standard assumptions for stationary time series. I use the powerful tie-break bootstrap that, as the included simulation study implies, allows me to detect asymmetries with up to 25 series and the number of observations corresponding to one year of daily returns. Applying the procedure to US portfolio returns separately for each year shows that the amount of copula central asymmetry is time-varying and less present in the recent past. Asymmetry is more critical in portfolios based on size and less in portfolios based on book-to-market and momentum. In portfolios based on industry classification, asymmetry is present during market downturns, coherently with the financial contagion narrative. \end{abstract}

\noindent%
\vfill

\newpage
\spacingset{1.8} 
\section{Introduction}\label{Intro}

Asymmetric dependence is now considered an unavoidable characteristic of financial returns. Financial markets exhibit an increase in cross-sectional dependence during financial crises through different mechanisms collectively called financial contagion. This different probability of co-movements in the lower tail of the returns distribution generates an asymmetry in the dependence see, e.g., \cite{longin2001extreme}; \cite{ang2002asymmetric}; \cite{hong2007asymmetries}.
Market supervisors' financial stability considerations are highly influenced by dependence asymmetry and the systemic risk it entails.
Risk managers should include this characteristic in their probabilistic models and raise their capital buffer requirements to avoid losses in distress periods.
Asset managers should consider this feature by reallocating their assets on time to profit from arbitrage opportunities or avoid losses by making proper hedging decisions. 

Even in tranquil times, asset managers should include the skewness, co-skewness, and odd moments that asymmetric dependence generates in their portfolio optimization. \cite{Albuquerque2012} develops an equilibrium model of this relationship. For portfolio optimization with higher moments, see \cite{jondeau2007financial} and reference therein and expanding on the previous reference,  I additionally recall that the large class of mixed risk-averse agents \citep{Caballe1996, tsetlin2009} prefer the risk with a higher moment for odd moments and the risk with a lower moment for even moments. If odd moments are sizeable, portfolio allocation of mixed risk-averse agents could differ dramatically from the one obtained by agents using the mean-variance analysis.

Historically, several statistics were proposed for the detection of asymmetric dependence. The pioneering works of \cite{longin2001extreme}; \cite{ang2002asymmetric}; \cite{hong2007asymmetries}, used exceedance correlations. The positive (negative) exceedance correlation is the correlation conditional to the upper(lower) tail. A statistic for detecting asymmetric dependence is the difference between positive and negative exceedance correlation. Other similar statistics were differences between positive and negative exceedance covariances or exceedance betas \cite{hong2007asymmetries}. See \cite{chen2015} and reference therein for more recent work in this line of research.

The approach has three main drawbacks: it is conditionally linear and cannot capture non-linear dependencies in the tails. It is bivariate, including information only for pairs of assets. The statistic depends on the choice of the threshold(s) to distinguish the tails from the center of the distribution. 

A second approach, in the literature, is based on the difference between the upper and lower tail dependence functions\citep[see ][ and references therein]{Bormann2020}. Those measures are usually considered more robust non-linear alternatives to exceedance covariances and correlations. The approach is still bivariate or applied pairwise, controlling the family-wise error. It depends on a tuning parameter to select extreme observations. Adaptation to the time-series context is not firmly justified theoretically.

Based on a distributional distance, I propose a non-parametric test of central symmetry applied to Copula functions.
Central Symmetry is one possible generalization of univariate symmetry to higher dimensions. Elliptical distributions are a paradigmatic example of a centrally symmetric distribution. Under this symmetry, the upper and lower tails have the same probability.
Violation of the multivariate distribution's central symmetry could then imply dependence asymmetry. Unfortunately, testing for the multivariate distribution's central symmetry requires knowledge or estimation of the center of symmetry, a highly nontrivial task in the multivariate setting. 
For this reason, I test copula central symmetry, where the center is known as the unit hypercube's center. In addition, the central asymmetry of copula functions is more directly related to dependence asymmetry and not contaminated by marginal asymmetry. This symmetry also goes under the names copula radial symmetry or copula reflection symmetry.\footnote{see pages 64-65 of \cite{joe2014dependence} for a discussion on the different names.} Because I study the relationship between multivariate and copula symmetry, I adopt the copula central symmetry name in this paper.
Non-parametric tests of copula central symmetry based on distributional distances were pioneered by \cite{aki1993nonparametric} without even referring to copula functions. \cite{bouzebda2012test}, \cite{dehgani2013measures}, and \cite{genest2013tests} introduced the test in the copula literature in the bivariate case. See \cite{billio2021} for a complete list of references of the different tests of this property and an extensive simulation study in the i.i.d. case, comparing the main procedures.

Testing non parametrically copula central symmetry is a multivariate non-linear approach that does not require any tuning parameter nor the estimation of a center of symmetry. It is independent of marginal symmetry
and can be easily adapted to the context of weakly dependent data.

In this respect, a recent powerful tie-break bootstrap \cite{Seo2024}, valid also for time series and useful only in the copula setup, allows statistical power with a large number of series and a small number of observations. The basic idea at the core of the new bootstrap was developed in the context of a randomization test \cite{beare2020} for copulas. \cite{billio2022} investigated the outstanding performance of the randomization procedure of \cite{beare2020} for testing copula central symmetry in the i.i.d. case. Unfortunately, the randomization test was not easily portable to time series. In this paper, I show comparable performance improvements, in the case of time series, using the tie-break bootstrap of \cite{Seo2024} in testing copula central symmetry.

In the empirical application, I use the benchmark dataset of  US portfolio returns
available from Kenneth French. This dataset was used in \cite{ang2002asymmetric},  \cite{Hong2006},\cite{Bormann2020}, and several other papers in the dependence asymmetry literature. Differently from the previous studies, the power of our test allows me to test symmetry separately for each year in the sample, investigating the time-varying dimension of dependence asymmetry using non-overlapping windows.

The paper is structured as follows. Section \ref{sec:central} defines central symmetry, discusses the relationships between the different measures of dependence asymmetry. Section \ref{sec:test} introduces the test, the tie-break bootstrap, and studies the statistical power of the procedure by simulation. Section \ref{sec:empirical} apply the test of copula central symmetry to US equity portfolios based on different characteristics. Section \ref{sec:conclusions} concludes by summarizing our findings and outlining future research directions.

\section{Central Symmetry}\label{sec:central}
\begin{figure}[htb]
\adjustbox{scale=0.6,center}{
\begin{tikzcd}
	{\hbox{Marginal Symmetry}} &&&& {\hbox{Zero Odd Moments}} \\
	&&&& {\hbox{Univariate Symmetry of Portfolios}} \\
	& {+} & {\hbox{Central Symmetry}} \\
	&&&& {\hbox{Equal Upper and Lower Exceedance Correlations}} \\
	{\hbox{Copula Central Symmetry}} &&&& {\hbox{Equal Upper and Lower Tail Dependence Functions}}
	\arrow["{Theorem\,\ref{margcop}}"', from=1-1, to=3-2]
	\arrow[from=3-2, to=3-3]
	\arrow["{Corollary \ref{CS->MS}}"', from=3-3, to=1-1]
	\arrow["{Proposition\, \ref{oddmom}}", from=3-3, to=1-5]
	\arrow["{Proposition\, \ref{oddmom}}"', from=3-3, to=2-5]
	\arrow["{Proposition\, \ref{exced}}"{description}, from=3-3, to=4-5]
	\arrow["{Theorem\,\ref{margcop}}", from=3-3, to=5-1]
	\arrow["{Theorem\,\ref{margcop}}", from=5-1, to=3-2]
	\arrow["{Lemma \ref{taildep}}"{description}, from=5-1, to=5-5]
\end{tikzcd}}
\caption{Summary of the relationship between the different symmetry concepts and properties important in financial markets. Arrows are implications; $+$ means all the incoming implications must hold.  }
\label{fig:implications}
\end{figure}
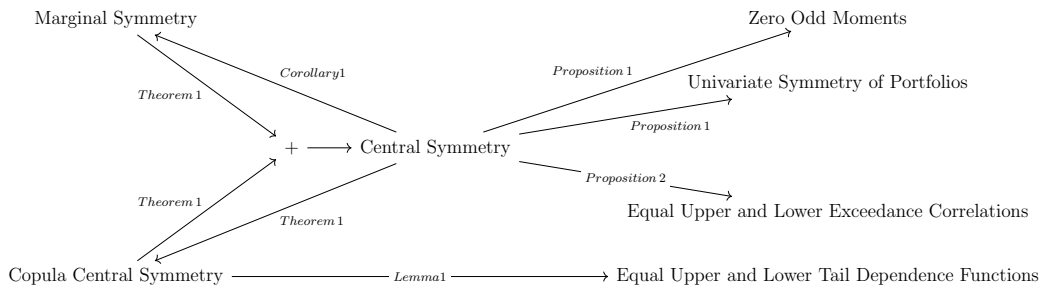

In this section, I introduce several multivariate symmetry concepts for random vectors and relate them to properties important in the financial context or used in dependence asymmetry measurement. I summarize these theoretical results in figure \ref{fig:implications}. 

I denote a random quantity by capital letters and deterministic quantities by lowercase letters.  ${\buildrel d \over =}$ denotes the equality in distribution.\\
 I start by considering univariate symmetry.
\begin{defi}[Univariate Symmetry]\label{US}
 A random variable $R$ is symmetric around a center  $c$ if $X=R-c$ has the same distribution of $Y=-X=c- R$ or $R-c{\buildrel d \over =}c- R$.   
\end{defi}

In particular, if I introduce the cumulative distribution function (CDF) of $X$, $F_{X}\left(x\right) = \mathbb{P}\left(X\leq x\right)$ and the survival function (SF), $\bar{F}_{X}\left(x\right) = \mathbb{P}\left(X> x\right)=1-F_{X}\left(x\right) $,  then, if the CDF is continous\footnote{ Otherwise \eqref{unisymm} holds only at the points of continuity
of the CDF}
\begin{eqnarray}\label{unisymm}
F_{Y}\left(x\right) = \mathbb{P}\left(Y\leq x\right)    = \mathbb{P}\left(-X\leq x\right)  = \mathbb{P}\left(X> -x\right)  =\bar{F}_{X}\left(-x\right)       
\end{eqnarray}
and  symmetry is valid if and only if the equality  \[F_{X}\left(x\right)=F_{Y}\left(x\right)=\bar{F}_{X}\left(-x\right),\]
holds for every $x\in\mathbb{R}$.
Several extensions of the symmetry property to a random vector are possible. In the following, bold symbols represent vectors, for example 
$\mathbf{R}=\left(R_1,\ldots,R_N\right)^{\prime}$, and inequalities with bold symbols represent joint inequalities, for example $\left\{\mathbf{R}\leq \mathbf{r} \right\}= \left\{\bigcap^{N}_{i=1} R_i \leq r_i\right\}$. 

The straightforward generalization of univariate symmetry is marginal symmetry :
\begin{defi}[Marginal Symmetry]\label{MS}
 A random vector $\mathbf{R}$ is marginally symmetric if every component is univariate symmetric.    
\end{defi}
Under this multivariate symmetry, I require that all the marginals are symmetric. This symmetry disregards the dependence among the components.  

Another possible generalization of univariate symmetry, taking dependence into account, is central symmetry:

\begin{defi} [Central Symmetry]\label{CS} A random vector $\mathbf{R}$ is centrally symmetric with respect to the center $\boldsymbol{c} = \left(c_{1},\ldots,c_{N}\right)$ if \begin{equation}\label{radsymeqdistr}\mathbf{R}-\boldsymbol{c}=\mathbf{X}{\buildrel d \over =}\boldsymbol{c}- \mathbf{R}=\mathbf{Y}. \end{equation}
\end{defi}

Let me define $\mathcal{N}= \left\{1,\ldots,N\right\}$ and  the multi-index $\mathcal{I}=\left\{i_1,\ldots,i_k\right\}\subseteq \mathcal{N}$  where $\left\vert\mathcal{I}\right\vert=K\leq N$. The subvector of $\mathbf{X}$ with components in  $\mathcal{I}$ is $\mathbf{X}_{\mathcal{I}} = \left(X_{i_1},\ldots,X_{i_k}\right)^{\prime}$.  The marginal CDF of $\mathbf{X}_{\mathcal{I}} $ is $F_{\mathbf{X}_{\mathcal{I}}} \left(\mathbf{x}_{\mathcal{I}}\right) = \mathbb{P}\left(\mathbf{X}_{\mathcal{I}}\leq \mathbf{x}_{\mathcal{I}}\right) $.
The multivariate SF $\bar{F}_{\mathbf{X}}\left( \mathbf{x}\right)=\mathbb{P}\left(\mathbf{X}> \mathbf{x} \right)$ and CDF $F_{\mathbf{X}}\left(\mathbf{x}\right)=\mathbb{P}\left(\mathbf{X}\leq \mathbf{x}\right)$, satisfy the following relationship
\begin{eqnarray}\label{multsurvcdf}
\bar{F}_{\mathbf{X}}\left(\mathbf{x}\right)= \displaystyle\sum^{N}_{K=0} \sum_{\begin{array}{c}\mathcal{I}\subseteq \mathcal{N }\\\left\vert\mathcal{I}\right\vert=K\end{array} } \left(-1\right)^{K} F_{X_{\mathcal{I}}} \left(\mathbf{x}_{\mathcal{I}}\right)
\end{eqnarray}
In analogy with the univariate case, central symmetry holds if and only if, the following property of continuous multivariate CDF and SF holds for every $\mathbf{x}\in \mathbb{R}^N$

\begin{equation}\label{rad}
    F_{\mathbf{X}}\left(\mathbf{x}\right)=F_{\mathbf{Y}}\left(\mathbf{x}\right)=\bar{F}_{X}\left(-\mathbf{x}\right),
\end{equation}

The next result shows the consequences of the central symmetry hypothesis, impacting financial markets.

\begin{prop}\label{oddmom}
If $\mathbf{R}$ is centrally symmetric, 
\begin{itemize}
    \item[i)] $\mathbb{E}\left[R_i\right]=c_i,$  $i=1,\ldots,N$
    \item[ii)] Odd moments are zero when they exist.
    \item[iii)] Every linear combination of $\mathbf{R}$ has a symmetric distribution. 
\end{itemize}    
\end{prop}

The proposition implies that under central symmetry, I can neglect skewness, co-skewness, and higher-order odd moments in portfolio optimization and that every portfolio built from a centrally symmetric $\mathbf{R}$ has a symmetric distribution.   Proposition \ref{oddmom} has the following important corollary as a subcase of iii):
\begin{coro}\label{CS->MS}
Central Symmetry implies Marginal Symmetry.    
\end{coro}
This result was already stated in the bivariate case in \cite{nelsen1993some}.

The literature on financial contagion focuses on dependence asymmetry in the tails conditioning on exceedances. Following  \cite{ang2002asymmetric} and \cite{hong2007asymmetries} the  positive (negative) exceedance at level $a$ of two random variable $R_1,R_2$ , given their standardization $\tilde{R}_i = \dfrac{R_1-\mu_i}{\sigma_i}$, $i=1,2$ is the event $\left\{\tilde{R}_1> a \right\}\cup \left\{\tilde{R}_2> a \right\} $,
$ \left(\left\{\tilde{R}_1<- a \right\}\cup \left\{\tilde{R}_2< - a \right\} \right)$.The means and standard deviations  conditional to the exceedance at level $a$ for $i=1,2$ are
\begin{eqnarray*}\nonumber
\mu^{+}_i\left(a\right) &=& \mathbb{E}\left(  \tilde{R}_i\right. \left\vert \tilde{R}_1> a ,\tilde{R}_2> a\right) , \mu^{-}_i\left(a\right) = \mathbb{E}\left(  \tilde{R}_i\right. \left\vert \tilde{R}_1< -a ,\tilde{R}_2< -a\right)\\
\sigma^{+}_i\left(a\right) &=& \sqrt{\mathbb{E}\left(  \left(\tilde{R}_i- \mu^{+}_i\left(a\right)\right)^2\right. \left\vert \tilde{R}_1> a ,\tilde{R}_2> a\right)} \\ \sigma^{-}_i\left(a\right) &=& \sqrt{\mathbb{E}\left(  \left(\tilde{R}_i- \mu^{-}_i\left(a\right)\right)^2 \right. \left\vert \tilde{R}_1< -a ,\tilde{R}_2< -a\right)}
\end{eqnarray*}
the covariances at exceedance at level $a$ are:
\begin{eqnarray*}
\sigma^{+}_{1,2}\left(a\right)&=& \mathbb{E}\left( \left(\tilde{R}_1- \mu^{+}_1\left(a\right)\right) \left(\tilde{R}_2- \mu^{+}_2\left(a\right)\right) \right. \left\vert \tilde{R}_1> a ,\tilde{R}_2> a\right)\\
\sigma^{-}_{1,2}\left(a\right)&=& \mathbb{E}\left( \left(\tilde{R}_1- \mu^{-}_1\left(a\right)\right) \left(\tilde{R}_2- \mu^{-}_2\left(a\right)\right) \right. \left\vert \tilde{R}_1< -a ,\tilde{R}_2< -a\right)
\end{eqnarray*}
Finally, correlations at exceedance at level $a$ are defined by the following equations
\begin{eqnarray*}
\rho^{+}\left(a\right) = \dfrac{\sigma^{+}_{1,2}\left(a\right)}{\sigma^{+}_{1}\left(a\right)\sigma^{+}_{2}\left(a\right)}, \rho^{-}\left(a\right) = \dfrac{\sigma^{-}_{1,2}\left(a\right)}{\sigma^{-}_{2}\left(a\right)\sigma^{-}_{1}\left(a\right)}
\end{eqnarray*}

The following proposition studies the effect of central symmetry on exceedance covariance and correlations 
\begin{prop}\label{exced}
If $\mathbf{R}=\left(R_1,R_2\right)^{\prime}$ is centrally symmetric, 
\begin{eqnarray*}
\mu^{+}_i\left(a\right)&=&-\mu^{-}_i\left(a\right), \quad i=1,2\\
\sigma^{+}_i\left(a\right)&=& \sigma^{-}_i\left(a\right), \quad i=1,2\\   
\sigma^{+}_{1,2}\left(a\right)&=& \sigma^{-}_{1,2}\left(a\right)\\
\rho^{+}\left(a\right) &=& \rho^{-}\left(a\right)
\end{eqnarray*}  
\end{prop}

The proposition holds for each pair of components in case $\mathbf{R}$ is $N$-dimensional and centrally symmetric.\\

 It is helpful to differentiate the role of marginal random variables from the contributions of their dependence structure in a different way. This task can be accomplished using notions from copula theory. I provide only the results and definitions needed in the following, but the interested reader could refer to \cite{nelsen2007introduction}, \cite{joe2014dependence} and \cite{durante2015principles}.

Let $F_{X_i}\left(x_i\right)$,  be the marginal cumulative distribution function (CDFs) of $X_i$, the $i$-th component of  $\mathbf{X}$. The component-wise probability integral transform (PIT) applied to $\mathbf{X}$ leads to a random vector $\mathbf{U}$ with components distributed uniformly on the unit interval:

 \begin{equation}\label{U}
 \mathbf{U}=\left(U_1,\ldots,U_N\right)=\left(F_{X_1}\left(X_1\right),\ldots,F_{X_N}\left(X_N\right)\right).    
 \end{equation}

Following \cite{1959fonctions}, if the marginals are continous, the joint CDF of $\mathbf{X}$ equivalent to:
\begin{eqnarray*}
F_{\mathbf{X}}\left(\mathbf{x}\right)= C_{\mathbf{U}}\left(F_{X_1}\left(x_1\right),\ldots,F_{X_N}\left(x_N\right)\right),
\end{eqnarray*}
 $C_{\mathbf{U}}$  is the copula corresponding to  $F_{\mathbf{X}}$ and represents the joint CDF of $\mathbf{U}$. An equivalent strategy applies to the marginal survival functions of $\mathbf{X}$, $\bar{F}_{X_i}\left(x_i\right)$, $ i\in \mathcal{N}$. At each $\mathbf{x}\in \mathbb{R}^N$, I can write the  joint survival function of $\mathbf{X}$:
\begin{eqnarray*}
\bar{F}_{\mathbf{X}}\left(\mathbf{x}\right)= \bar{C}_{\mathbf{U}}\left(\bar{F}_{X_1}\left(x_1\right),\ldots,\bar{F}_{X_N}\left(x_N\right)\right).
\end{eqnarray*}
Then, an alternative expression for the joint CDF of $\mathbf{Y}$ is:
\[ F_{\mathbf{Y}}\left(\mathbf{x}\right) =\bar{F}_{X}\left(-\mathbf{x}\right)=\bar{C}_{\mathbf{U}}\left(\bar{F}_{X_1}\left(-x_1\right),\ldots,\bar{F}_{X_N}\left(-x_N\right)\right)\]

and 

\[\left(F_{Y_1}\left(Y_1\right),\ldots,F_{Y_N}\left(Y_N\right)\right)=\left(\bar{F}_{X_1}\left(-\left(-X_1\right)\right),\ldots,\bar{F}_{X_N}\left(-\left(-X_N\right)\right)
\right)=\mathbf{1}_N- \mathbf{U},
\]
where  $\mathbf{1}_N$ be a $N$-dimensional vector with all components equal to one. 
The survival copula $\bar{C}_{\mathbf{U}}$ represents the joint CDF (not the joint SF) of  $\mathbf{1}_N-\mathbf{U}$. The following definition of copula central symmetry follows:\\
\begin{defi}[Copula Central Symmetry]\label{CCS}
  A random vector $\mathbf{R}$ is Copula Centrally Symmetric if the corresponding $\mathbf{U}$ from \eqref{U}, satisfies $\mathbf{U} {\buildrel d \over =} \mathbf{1}_N- \mathbf{U}$.  
\end{defi}

To understand the relationship between central symmetry, marginal symmetry, and copula central symmetry, I define the symmetrization and the antisymmetrization of  $C_{\mathbf{U}}$
\begin{eqnarray}
\Delta_{\mathbf{U}}\left(\mathbf{u}\right)&=& \dfrac{1}{2} \left( C_{\mathbf{U}}\left(\mathbf{u}\right) -\bar{C}_{\mathbf{U}}\left(\mathbf{u}\right)\right)\\
C_{\mathbf{U}^{S}}\left(\mathbf{u}\right)&=& \dfrac{1}{2} \left( C_{\mathbf{U}}\left(\mathbf{u}\right) +\bar{C}_{\mathbf{U}}\left(\mathbf{u}\right)\right)  
\end{eqnarray}
I remark that $C_{\mathbf{U}^{S}}$, being a convex combination of copula functions is a copula. It is the CDF of 
\begin{eqnarray}
\mathbf{U}^{S}= \left\{\begin{array}{cc}
  \mathbf{U}   & \hbox{ with probability } \frac{1}{2} \\&\\
    \mathbf{1}_N-  \mathbf{U}   & \hbox{ with probability } \frac{1}{2} .
\end{array}\right.
\end{eqnarray}

The following Theorem disentangles the contribution coming from marginal and dependence asymmetry.

\begin{theorem}
    \label{margcop}
\begin{eqnarray}  \nonumber
  F_{\mathbf{X}}\left(\mathbf{x}\right)&-&\bar{F}_{X}\left(-\mathbf{x}\right)=   
  \\\label{radmarg} C_{\mathbf{U}^{S}}\left(F_{X_1}\left(x_1\right),\ldots, F_{X_N}\left(x_N\right)\right) &-&C_{\mathbf{U}^{S}}\left(\bar{F}_{X_1}\left(-x_1\right),\ldots,\bar{F}_{X_N}\left(-x_N\right)\right)
  \\ \label{radcop}
  + \Delta_{\mathbf{U}} \left(F_{X_1}\left(x_1\right),\ldots, F_{X_N}\left(x_N\right)\right) &+& \Delta_{\mathbf{U}} \left(\bar{F}_{X_1}\left(-x_1\right),\ldots,\bar{F}_{X_N}\left(-x_N\right)\right).
  \end{eqnarray}
  In addition,
  \begin{enumerate}
      \item   \eqref{radmarg}  is zero for all $\mathbf{x}\in \mathbb{R}^N$ if and only if marginal symmetry holds.
      \item 
      \eqref{radcop} is zero for every $\mathbf{x}\in \mathbb{R}^N$ if and only if copula central symmetry holds.
       \item $\mathbf{R}$ is centrally symmetric if and only if it is jointly marginally symmetric and copula centrally symmetric.
  \end{enumerate}
  
\end{theorem}

 The line \eqref{radmarg} represents the contribution to central asymmetry coming from the marginals embedded in the symmetrization of the dependence structure. The line \eqref{radcop} adds the contributions from the dependence structure. Point 3 of Theorem \ref{margcop} is the generalization to the multivariate case of Theorem 3.2 in \cite{nelsen1993some}. The decomposition is new even in the bivariate case and shows a compensation between marginal asymmetry and dependence asymmetry if they have opposite signs.
The different sign of those contributions in financial markets is empirically documented and theoretically motivated in \cite{Albuquerque2012}. In particular, the latter paper shows positive skewness and negative co-skewness. The Theorem and the empirical findings in \cite{Albuquerque2012} imply that negative dependence asymmetry, usually associated with financial contagion, could be masked by positive marginal symmetry if we use measures based on the multivariate distribution as exceedance correlations and covariances. In this work, instead, I focus on the asymmetry coming from the dependence structure and the null hypothesis will be of copula central symmetry.
 Copula central asymmetry has consequences for other measures of asymmetric dependence. Those measures are based on the difference between the upper and lower tail dependence functions\citep[see ][ and references therein]{Bormann2020}.   

The lower tail dependence function of $\mathbf{1}_N- \mathbf{U}$ is equal to the upper dependence function of $\mathbf{U}$ \citep[see, for example, section 2.18 in ][]{joe2014dependence} leading to the following result that I state as a Lemma: 
\begin{lemma}\label{taildep}
Copula central symmetry implies that upper and lower tail dependence functions are equal.   
\end{lemma}
 Analogously more refined upper and lower tail orders \cite{Hua2011} and upper and lower directional \citep{joe2010} and directional tail-weighted dependence measures \citep{Li2024} are equal under central symmetry. Copula central Symmetry is not impacted by marginal symmetry as central symmetry and implies the nullity of measures based on upper and lower tail dependence. In addition, as remarked in the introduction, it is a genuinely non-linear and multivariate property. On theoretical grounds, testing copula central symmetry appears to be the best procedure to detect dependence asymmetry. 

\section{ Testing Time Series for Copula Central Symmetry}\label{sec:test}
In this section, I introduce the test statistic and the tie-break bootstrap for testing the null hypothesis of copula central symmetry
\[\mathcal{H}_0 : \mathbf{U} {\buildrel d \over =} \mathbf{1}_N- \mathbf{U}.\]

or equivalently using the copula and the survival copula 

\[\mathcal{H}_0 : C_{\mathbf{U}}\left(\mathbf{u}\right) =\bar{C}_{\mathbf{U}}\left(\mathbf{u}\right)\]

 An informative test statistic for the null hypothesis $\mathcal{H}_0$, could be a functional of $\Delta_{\mathbf{U}}\left(\mathbf{u}\right)$. 
 A non-parametric consistent estimation of $\Delta_{\mathbf{U}}$ can use the Empirical Copula. Let us consider an independent sample of size $T$ from $N$-dimensional random vector $\mathbf{X}$, $\left\{\left\{X_{ti}\right\}^{T}_{t=1}\right\}^{N}_{i=1}\equiv\left\{\mathbf{X}_{t}\right\}^{T}_{t=1}$ be . I denote the set $A$ indicator as $\mathbb{I}\left(A\right)$.
In addition, I define the normalized ranks  $U_{T,ti}=\dfrac{1}{T+1}\sum^T_{s=1}\mathbb{I}\left(X_{s i}\leq X_{t i}\right)$, $t=1,\ldots,T$ and $i\in\mathcal{N}$ .
With those definitions, the empirical copula and the empirical survival copula are:
\begin{eqnarray}\nonumber
C_{T}\left(\mathbf{u}\right)=\displaystyle \dfrac{1}{T}\sum^T_{t=1}\mathbb{I}\left(\mathbf{U}_{T,t}\leq \mathbf{u}\right),\quad
\bar{C}_{T}\left(\mathbf{u}\right)  = \displaystyle \dfrac{1}{T}\sum^T_{t=1}\mathbb{I}\left(\mathbf{1}_N-\mathbf{U}_{T,t}\leq \mathbf{u}\right).
\end{eqnarray}

 Then, the antisymmetrization of  the empirical copula is
\begin{eqnarray}
\Delta_{T}\left(\mathbf{u}\right)&=& \dfrac{1}{2} \left( C_{T}\left(\mathbf{u}\right) -\bar{C}_{T}\left(\mathbf{u}\right)\right)
\end{eqnarray}

 An application of the functional delta method \cite{van1996weak} on results for central limit theorem (CTL) of the multivariate empirical process for strongly mixing data with mixing coefficient $\alpha_n= o\left(n^{-a}\right)$ for some $a>0$, that can be found in \cite{rio1999theorie}, allow \cite{bucher2013consistent} to obtain the  weak convergence result for the empirical copula process under the following non-restrictive assumptions on copula derivatives

 \begin{hyp}\label{segersder1} For each $j\in \left\{1, \dots , d\right\}$, the jth first-order partial derivative
$\dfrac{\partial C}{\partial u_j}$
exists and is continuous on the set $V_{d,j}:= \left\{u \in \left[0, 1\right]^d : 0 < u_j < 1\right\}$,
\end{hyp}

 Under assumption \textbf{A} \ref{segersder1}, for strongly mixing data with mixing coefficient $\alpha_n= o\left(n^{-a}\right)$ for some $a>0$, the empirical copula process  $\mathbb{C}_T=\sqrt{T}\left(C_{T}\left(\mathbf{u}\right)-C_{\mathbf{U}}\left(\mathbf{u}\right)\right)$
converge weakly, in the Hoffman-Jorgensen sense, in $\ell^{\infty}\left(\left[0,1\right]^N\right)$ the space of bounded function on the $N$-dimensional unit hypercube:
\begin{eqnarray}\label{weakcop}
\mathbb{C}_T\leadsto \mathbb{C}=\mathbb{B}_{\mathbf{U}}\left(\mathbf{u}\right)- \sum^{N}_{i=1}\dfrac{\partial C_{\mathbf{U}}\left(\mathbf{u}\right)}{\partial u_i}\mathbb{B}_{i,\mathbf{U}}\left(u_i\right),
\end{eqnarray}
where $\mathbb{B}_{_{\mathbf{U}}}$ is a d-dimensional Brownian sheet with covariance function

\begin{eqnarray}
\mathbb{C}\mathrm{ov}\left(\mathbb{B}_{\mathbf{U}}\left(\mathbf{u}\right),\mathbb{B}_{\mathbf{U}}\left(\mathbf{v}\right)\right)=\displaystyle\sum^{\infty}_{t=-\infty}\mathbb{C}\mathrm{ov}\left(\mathbb{I}\left(\mathbf{U}_0\leq\mathbf{u}\right),\mathbb{I}\left(\mathbf{U}_t\leq\mathbf{v}\right) \right)
\end{eqnarray}

In the following proposition, I derive the weak convergence results for the empirical survival copula process for strongly mixing data, under assumption \textbf{A} \ref{segersder1}.

\begin{prop}\label{empsurv}
 Suppose Conditions \textbf{A\ref{segersder1}} hold and the strongly mixing coefficients $\alpha_n$ of the sample are such that $\alpha_n= o\left(n^{-a}\right)$ for some $a>0$ . Then the empirical survival copula process $\mathbb{\bar{C}}_n=\sqrt{n}\left(\bar{C}_{n}\left(\mathbf{u}\right)-\bar{C}\left(\mathbf{u}\right)\right)$ weakly converges towards a Gaussian field  $\mathbb{\bar{C}}$ 
 \begin{eqnarray}\label{weaksurv}
\mathbb{\bar{C}}_T\leadsto \bar{\mathbb{C}}=\bar{\mathbb{B}}_{\mathbf{U}}\left(\mathbf{u}\right)- \sum^{N}_{i=1}\dfrac{\partial \bar{C}_{\mathbf{U}}\left(\mathbf{u}\right)}{\partial u_i}\bar{\mathbb{B}}_{i,\mathbf{U}}\left(u_i\right),
\end{eqnarray}
where $\bar{\mathbb{B}}_{\mathbf{U}}$ is a d-dimensional Brownian sheet with covariance function

\begin{eqnarray}\label{survcov}
\mathbb{C}\mathrm{ov}\left(\bar{\mathbb{B}}_{\mathbf{U}}\left(\mathbf{u}\right),\bar{\mathbb{B}}_{\mathbf{U}}\left(\mathbf{v}\right)\right)=\displaystyle\sum^{\infty}_{t=-\infty}\mathbb{C}\mathrm{ov}\left(\mathbb{I}\left(\mathbf{1}_N-\mathbf{U}_0\leq\mathbf{u}\right),\mathbb{I}\left(\mathbf{1}_N-\mathbf{U}_t\leq\mathbf{v}\right) \right)
\end{eqnarray}
\end{prop}

Several measures were proposed to test for copula central symmetry ( see \cite{billio2021} and references therein). I focus on  a Cram\'er–von Mises statistic under the random measure generated by the empirical copula:
\begin{eqnarray}\label{statistic}
S_{T}\left(\mathbf{U}_T\right)= \displaystyle\int_{\left(0,1\right]^d}\left(2\Delta_T\right)^2 dC_T = \dfrac{1}{T} \sum^{T}_{t=1} \left(2\Delta_{T}\left(\mathbf{U}_{T,t}\right)\right)^2.
\end{eqnarray}
The 2-dimensional version of this statistic was introduced in \cite{bouzebda2012test} and investigated further in \cite{dehgani2013measures} and \cite{genest2013tests}. This measure is one of the best performing for $d > 2$ in the i.i.d. case \citep{billio2021} and studied in the high dimensional case, in the context of a randomization test,  in \cite{billio2022}. I derive the asymptotic convergence in the time series setting in the following proposition.
\begin{prop}\label{statlim}
 Suppose Conditions \textbf{A\ref{segersder1}} hold and the strongly mixing coefficients $\alpha_n$ of the sample are such that $\alpha_T= o\left(T^{-a}\right)$ for some $a>0$ . Then Under Copula central symmetry $TS_{T}\left(\mathbf{U}_T\right)$  weakly converges towards $\mathbb{S}=\displaystyle\int_{\left(0,1\right]^d}\left(\mathbb{C}-\bar{\mathbb{C}}\right)^2 dC$ .
\end{prop}

\subsection{Tie Break Dependent Bootstrap}

 In this subsection, I describe the tie-break bootstrap procedure for time series introduced in \cite{Seo2024} and derive the asymptotic behavior of the bootstrapped version of our test statistic. 

\cite{Seo2024} proposes a bootstrap procedure improving finite sample performance by breaking ties induced by the block bootstrap on the bootstrapped normalized ranks. The tie-break bootstrap procedure for time series consists of the following steps:
\begin{enumerate}
    \item Divide $\mathbf{X}$ in $T-l_T +1$ blocks of $l_T$ consecutive observations\\
   $\left\{\mathbf{X}_1,\ldots, \mathbf{X}_{l_T}\right\}$, $\left\{\mathbf{X}_{l_T +1},\ldots, \mathbf{X}_{2l_T}\right\}, \ldots, \left\{\mathbf{X}_{T-l_T +1},\ldots, \mathbf{X}_{T}\right\}$.
   \item A  bootstrap sample $\mathbf{X}^{*}$ is obtained  by  sampling with replacement $T-l_T +1$ times from the block set.   
   \item  Compute the bootstrapped normalized ranks $U^{*}_{T,ti}=\dfrac{1}{T+1}\sum^T_{s=1}\mathbb{I}\left(X^{*}_{s i}\leq X^{*}_{t i}\right)$
    \item  Draw $T$ independent standard uniform random variables $\eta_1,\ldots,\eta_T$  and set , for $t=1,\ldots,T$
    \[\mathbf{U}^{*,\eta}_{T,t}=\mathbf{U}^{*}_{T,t}-n^{-1}\mathbf{1}_N \eta_t\].
    \item  For $t=1,\ldots,T$ and $i=1,\ldots,N$ compute:
    \[V^{*,\eta}_{T,ti}=\dfrac{1}{T}\sum^T_{s=1}\mathbb{I}\left(U^{*,\eta}_{T,si}\leq U^{*,\eta}_{T,ti}\right)\]
    \item Compute
    \begin{eqnarray*}
       C^{*,\eta}_{T}\left(\mathbf{u}\right)&=&\displaystyle \dfrac{1}{T}\sum^T_{t=1}\mathbb{I}\left(\mathbf{V}^{*,\eta}_{T,t}\leq \mathbf{u}\right)\\
\bar{C}^{*,\eta}_{T}\left(\mathbf{u}\right)  &= &\displaystyle \dfrac{1}{T}\sum^T_{t=1}\mathbb{I}\left(\mathbf{1}_N-\mathbf{V}^{*,\eta}_{T,t}\leq \mathbf{u}\right)\\
       S^{*,\left[1\right]}_{T}\left(\mathbf{V}^{*,\eta}\right)&=& \dfrac{1}{T} \sum^{T}_{t=1} \left(C^{*,\eta}_T\left(\mathbf{V}^{*,\eta}_{T,t}\right)-\bar{C}^{*,\eta}_T\left(\mathbf{V}^{*,\eta}_{T,t}\right)\right)^2\\
    \end{eqnarray*}
 
    \item Repeat steps 1-4 for a large number of times $M$ and compute the approximate p-value
    \begin{eqnarray}\label{randpval}
     \hat{P}&=&\dfrac{1}{M}\displaystyle\sum^{M}_{m=1} \mathbb{I}\left(S^{*,\left[m\right]}_{T}>S_{T}\right).
     \end{eqnarray}
\end{enumerate}

The procedure is valid under the following assumption on the data generating process and choice of $l_{T}$.

\begin{hyp}\cite{bucher2013consistent,Seo2024}\label{timeboot} 
\begin{itemize}
\item[(i)] $\mathbf{X}$ is drawn from a strictly stationary process with strong mixing and mixing coefficients $\alpha_T$ satisfying $\sum^{\infty}_{t=1}\left(T+1\right)^{16\left(d+1\right)}\sqrt{\alpha_T}<\infty$
\item[(ii)] $l_{T}= O\left(T^{1/2-k}\right)$ for some  $k$ s.t. $0<k<1/2$
\end{itemize}
\end{hyp}

Theorem 3.1 \cite{Seo2024} derive the following result under \textbf{A} \ref{segersder1} and \textbf{A} \ref{timeboot} 
\begin{equation}\label{tiebreakprocess}
\mathbb{C}^{*,\eta}_{T} =\sqrt{T}\left(   C^{*,\eta}_{T}- C_{T}\right) \overset{\mathbb{P}}{\underset{ \mathbf{X}}{\,\leadsto}\,} \mathbb{C}
\end{equation}
where the result is valid in $\ell^{\infty}\left(\left[0,1\right]^d\right)$ and $\overset{\mathbb{P}}{\underset{ \mathbf{X}}{\,\leadsto}\,}$ represents weak convergence conditional to the data in probability.
The following proposition derives the analogous result for the Tie Break survival empirical copula process converge weakly  conditional to the data in probability in $\ell^{\infty}\left(\left[0,1\right]^d\right)$ to the following limit 
\begin{prop}\label{tie-breaksurv}
under \textbf{A} \ref{segersder1} and \textbf{A} \ref{timeboot} , the Tie-Break bootstrap survival empirical copula process $\bar{\mathbb{C}}_T$ converge weakly conditional to the data in probability in $\ell^{\infty}\left(\left[0,1\right]^d\right)$ to the following limit:
\begin{equation}
\bar{\mathbb{C}}^{*,\eta}_{T} =\sqrt{T}\left(   \bar{C}^{*,\eta}_{T}- \bar{C}_{T}\right) \overset{\mathbb{P}}{\underset{ \mathbf{X}}{\,\leadsto}\,} \bar{\mathbb{C}}
\end{equation}
\end{prop}

The validity of the approximate P-value in equation \eqref{randpval}  comes from the following proposition whose proof is analogous to the proof of proposition \ref{statlim}.
\begin{prop}\label{limittest}
Under \textbf{A} \ref{segersder1} and \textbf{A} \ref{timeboot} 
 and the null of copula central symmetry $C= \bar{C}$  when $T\rightarrow\infty$, 
\begin{eqnarray*}
\left(TS^{\left[1\right]}_{T},\ldots,TS^{\left[M\right]}_{T}\right)&\overset{P}{\underset{\mathbf{X}}{\leadsto}} & \left( \mathbb{S}^{\left[1\right]},\ldots,\mathbb{S}^{\left[M\right]}\right),
\end{eqnarray*}
where $\mathbb{S}^{\left[1\right]},\ldots,\mathbb{S}^{\left[M\right]}$ are independent copies of $\mathbb{S}$.
\end{prop}

\subsection{Statistical Power}\label{power}
\begin{figure}[h]\label{fig: power}
\centering
\includegraphics[scale=0.5]{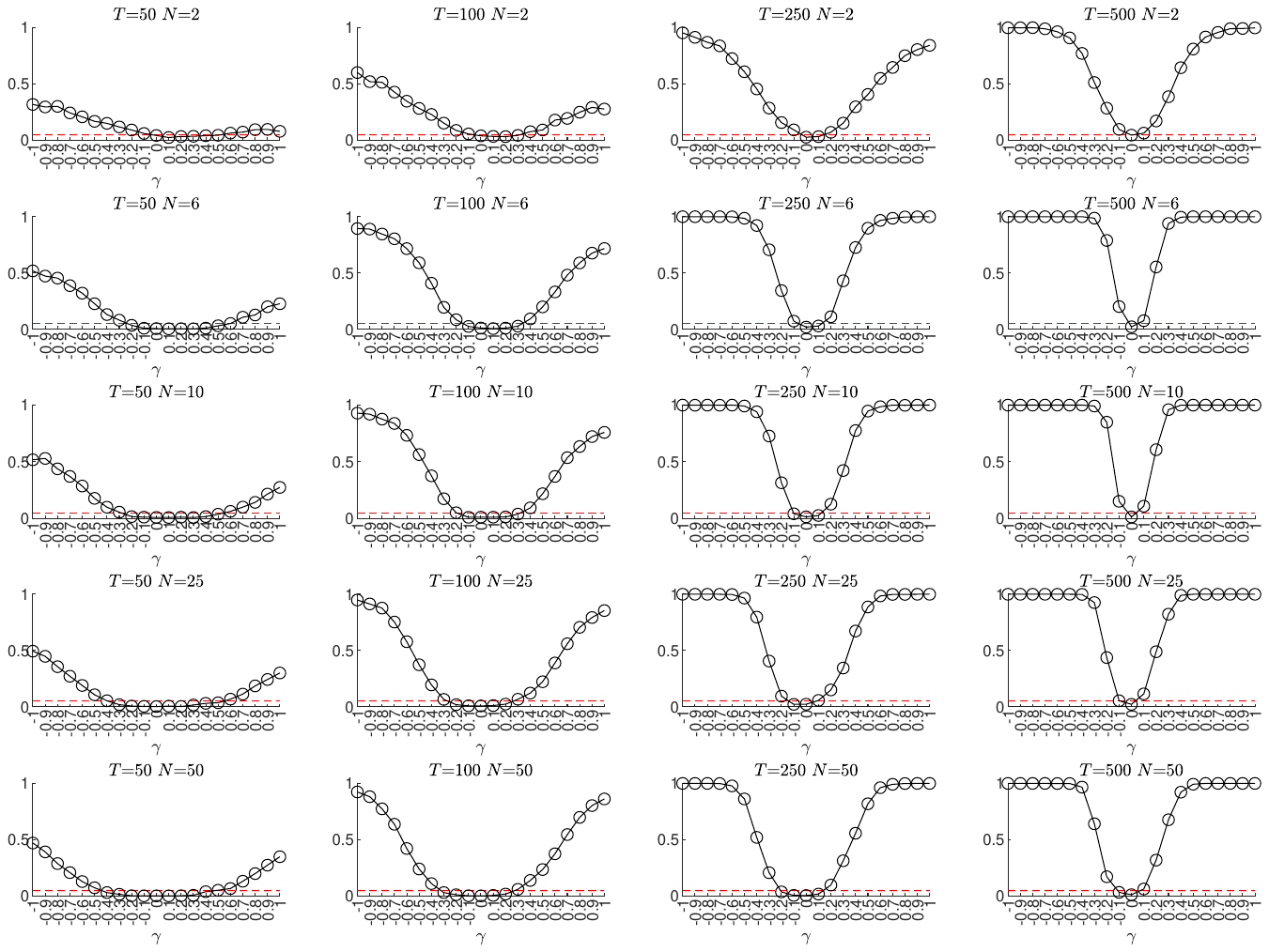}
\caption{Percentages of rejection at 5\% significance level in black, as estimated from 1000 replicates of the Tie-Break Bootstrap, for the tests based on $S_n$ under the Skew-t score driven factor model and different levels of asymmetry $\gamma$. The dashed red line is the nominal level of 5\%.}
\end{figure}

This section uses simulations to study the finite sample properties of the proposed multivariate copula central symmetry test. In all the experiments performed, the number of bootstrap or randomization replicates is M = 250, and the estimated rejection probabilities are computed using 1000 Monte Carlo independent replicates. Each table in the section presents a different number of observations $T\in\{50,100,250,500\}$, dimension of the random vector $N\in\{2,6,10,25,50\}$.
To study the power of the tests based on $S_{n}$ in the time series large dimensional context, I use a recent asymmetric generalization of a score-driven copula factor model introduced in \cite{ohpatton} with factors distributed according to a Skew-t distribution with common asymmetry parameter $\gamma\in\left[-1,1\right]$. I vary $\gamma$ in steps of $0.1$ and choose a number of groups equal to $2$ for $N\leq 10$ and $5$ in the other cases. The rest of the parameter values are from the simulation study in \cite{ohpatton}. Here and in the empirical application section the block length follows the heuristic proposed in \cite{bucher2013consistent}
The results are reported in figure \ref{fig: power}.   
The figure shows good statistical power of the test even in the case of small asymmetry and large dimensions if I consider more than 250 observations.  \cite{ohpatton} in their supplementary material report a value of $\gamma$ between -0.4  and  -0.2 for their main specifications. I report a satisfactory power level in comparable cases even with 25 series. On the contrary, considering 50 series $\gamma=-0.2$, the power level is below the significance level, even if it is satisfactory for higher values of $\gamma$.

\section{Empirical Evidence in US Stock Market}\label{sec:empirical}

In this section, I study the copula central asymmetry of equity portfolio returns composed of US stocks. I use a benchmark dataset publicly available thanks to Kenneth French in the data library section of his website\footnote{\url{http://mba.tuck.dartmouth.edu/pages/faculty/ken.french/data_library.html}}. 
 At first, I use the daily returns of five groups composed of 10 portfolios chosen by industry, deciles of market value, deciles of book-to-market ratio (value), and deciles of past performance(momentum)\footnote{A detailed description of the dataset is available on the Kenneth French website}.
  I consider data from 1 January 1927 to  31 December 2023 and test for central symmetry of the joint copula of the ten portfolios. 

\begin{figure}[h]
 \centering
\includegraphics[scale=0.5]{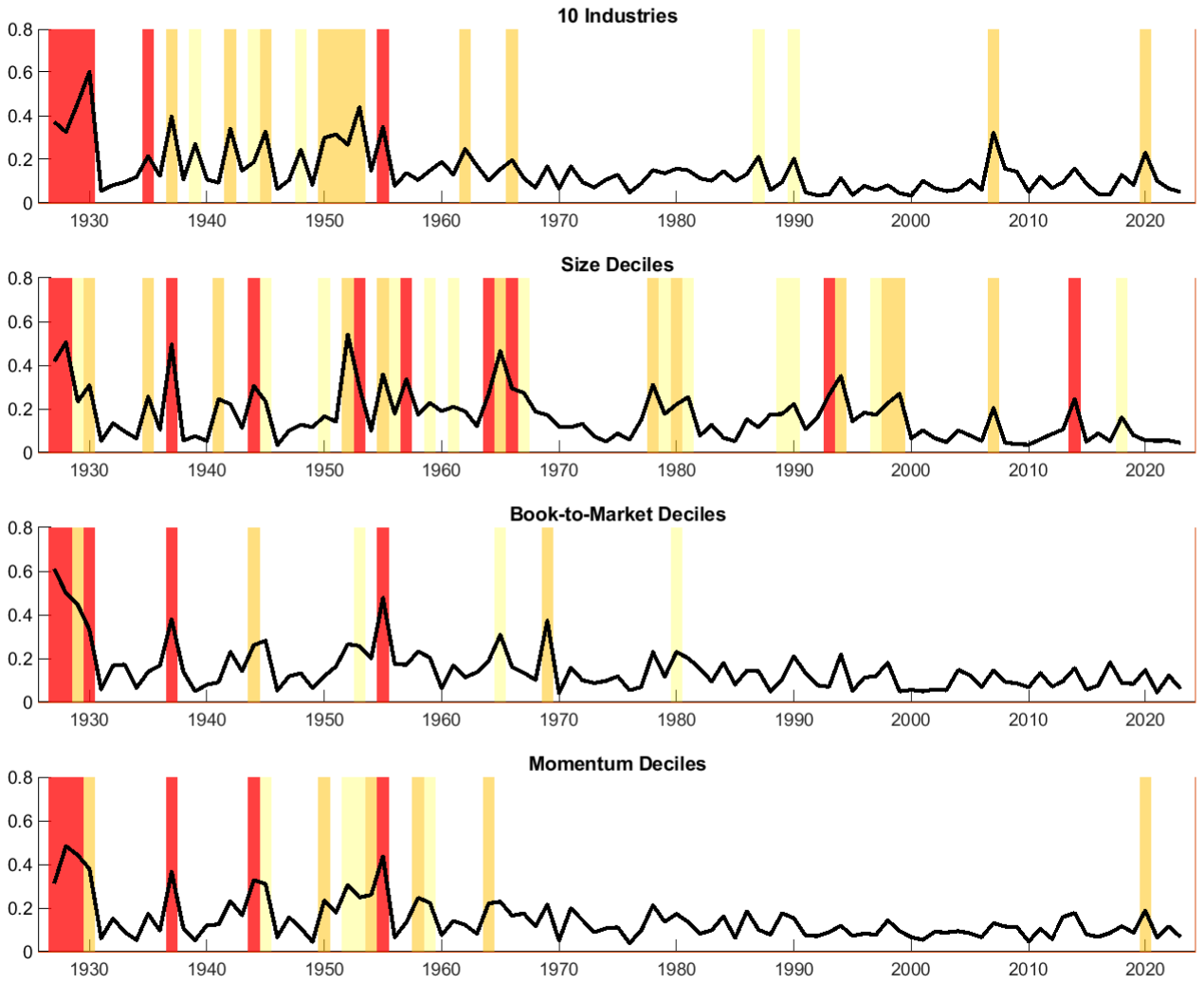}
\caption{Test Statistics(black line) and test of copula central symmetry for different types of portfolios with 10 components.
The procedure is performed separately for each year in the sample, controlling for the false discovery rate of all the years tested. Red bars are rejections at the nominal levels 1\%,
orange bars are rejections at the nominal level 5\% , and yellow bars are rejections at the nominal level 10\%  }
\label{fig:10}
\end{figure}

The statistical power outlined in the subsection \ref{power} allows me to test for copula central symmetry each year. The results are reported in figure \ref{fig:10}.
\begin{figure}[h]
\centering
\includegraphics[scale=0.5]{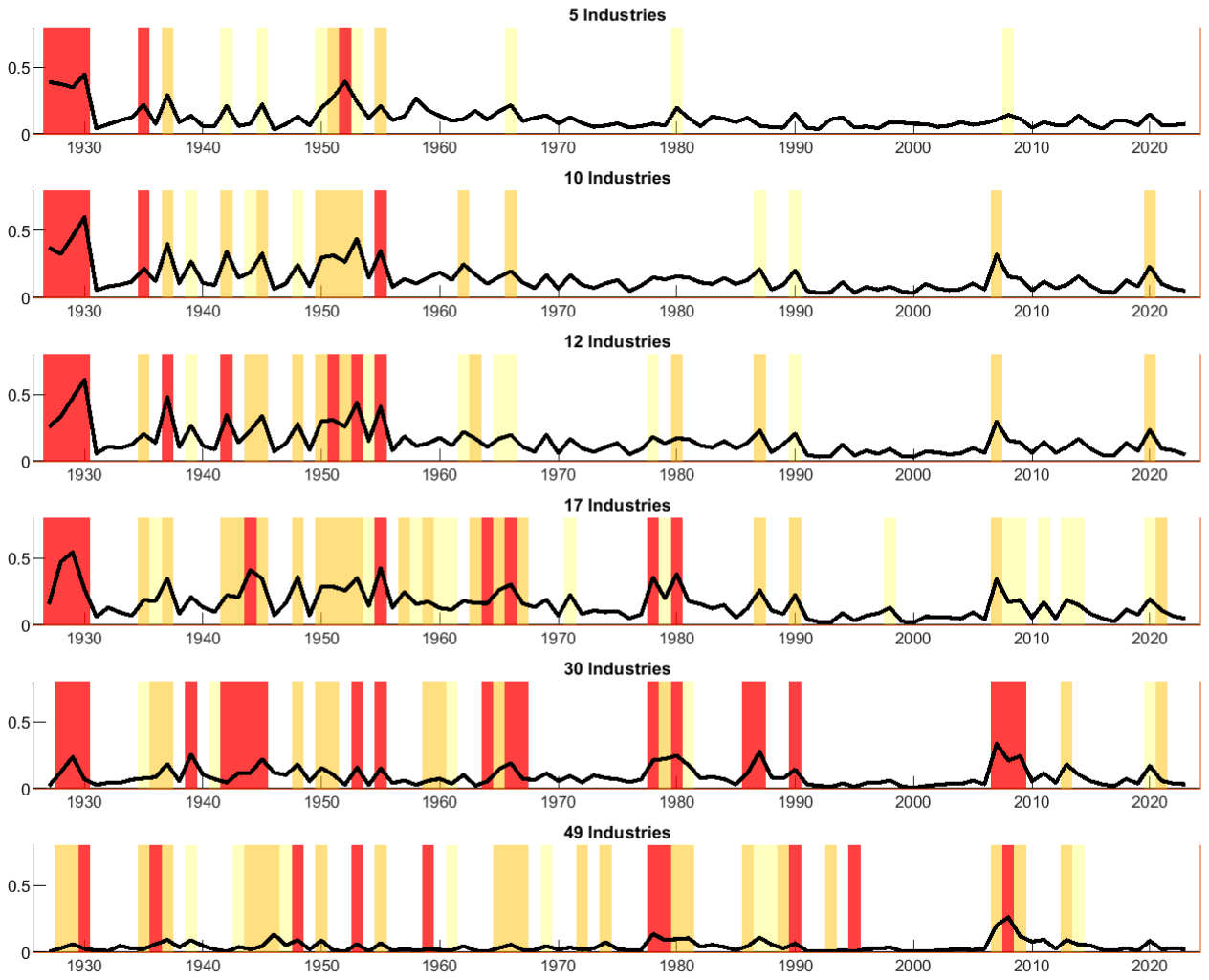}
\caption{Test Statistics(black line) and test of copula central symmetry for different granularity levels of industrial classification.
The procedure is performed separately for each year in the sample, controlling for the false discovery rate of all the years tested. Red bars are rejections at the nominal levels 1\%,
orange bars are rejections at the nominal level 5\% , and yellow bars are rejections at the nominal level 10\% }
\label{fig:sic}
\end{figure}
For each group of portfolios, I perform one test for each year, i.e., 97 tests. Since the number of tests is large, I control the whole testing procedure's false discovery rate (FDR) of each group of portfolios using the approach of \cite{FDR}.
I report results controlling for the FDR at the nominal levels 1\% (red), 5\% (orange), and 10\% (yellow). 
The amount of copula central asymmetry is time-varying for all five groups of portfolios and is less present in the second part of the sample. In particular, the 1929 crisis led to a strongly significant asymmetric event for all five groups, while the 2008 subprime crisis is relevant in only two of the four groups, and even in those cases, the magnitude of the statistics in the two years is not comparable. 
Asymmetry is more robust if we choose different sizes and weaker if we consider assets with different book-to-market or momentum. Strong asymmetry in size is consistent with the bivariate tests in \cite{ang2002asymmetric} and  \cite{Hong2006}. Weaker asymmetry in book-to-market is consistent with the bivariate analysis in \cite{Hong2006} results, while we have less asymmetry in momentum. In particular, asymmetry almost disappears for the latter two characteristics in the second part of the sample. 
 Using industry classification leads to less asymmetry than size but more asymmetry than book-to-market or momentum. In addition, asymmetry, in this case, is present during market downturns, consistently with the financial contagion narrative. This is particularly evident in the second part of the sample in which the Black Monday 1987, the subprime 2008, and the COVID 2020 are all labeled by the test as asymmetric years. I further explore this relationship between asymmetries and financial crisis with a detailed analysis based on industry classification at different levels of granularity in figure \ref{fig:sic}.
Increasing sector granularity seems to capture more asymmetry. The exception is the finest level of granularity of 49 industries, but, in this case, the analysis of statistical power of the subsection \ref{power} shows that the test is not entirely reliable with a comparable number of series. In addition, increasing granularity reduces differences in the first and the second part of the sample. For example, the peak of the statistic for 1929 becomes more and more comparable with the peak of 2008 if I increase the number of series considered. With 30 industries, the peaks are significant at the same confidence level and are comparable in magnitude. Those findings can be justified by the possibility that asymmetry develops in different parts of the economy during different times because the importance and riskiness of various industries change over time. Increasing the granularity of the industrial classification then appears more robust in capturing asymmetry.

\section{Conclusion}\label{sec:conclusions}
The detection of dependence asymmetry is relevant for portfolio optimization, market asymmetry, and financial stability. I focus on copula central asymmetry and explain, by new theoretical results, its relationship with other properties used by previous authors to measure dependence asymmetry. The results also imply that the detection of dependence asymmetry by some of these other measures can be obfuscated by marginal symmetry of the opposite sign. Testing copula central symmetry represents a multivariate non-linear approach that does not require tuning parameters or estimation of a center of symmetry. By construction, it is independent of marginal symmetry. I adapt to the time series context, a previously developed test based on distributional distance, using a novel, robust bootstrap framework introduced in \cite{Seo2024}. A simulation study with a state-of-the-art copula factor model DGP shows reliable statistical power with only 250 observations and a number of series less or equal to 25. The power of the test allows a yearly application of the procedure using daily returns data spanning almost a century for groups of portfolios based on different characteristics. I find that dependence asymmetry is a time-varying property and has been less relevant recently. Asymmetry is more present in portfolios based on size and less in portfolios based on book-to-market and momentum. In portfolios based on industry classification, asymmetry appears to be linked to market downturns in line with the financial contagion narrative. Increasing the granularity of the classification considered leads to a more robust detection of asymmetry due to the time-varying riskiness of different parts of the economy. The detection of dependence asymmetry analyzed in this paper could be extended in several directions. First, the test statistic cannot tell us the sign of the asymmetry, which would be essential given that a large class of utility functions implies a preference for positive asymmetry. Given a signed measure, it is possible to devise a portfolio optimization strategy that includes the measure. The measure introduced in \cite{Krupskii2016} has this characteristic but showed lower statistical performance using more than two series\citep{billio2021} in the i.i.d. case. The importance of this line of research for financial stability should aim instead to a different methodological advance. Given the time-varying nature of asymmetry, a structural break in this characteristic could be used as an early warning indicator for financial turmoil. The suggested change point detection analysis requires the extension of the theorems of this paper to the sequential empirical copula process \citep[see ][ and reference therein]{Bucher2014}.

\newpage
 \appendix

 \section{Proofs}
\subsection{Proof of Proposition \ref{oddmom}}
 \begin{proof}
i)-ii) I define $m_i\in\left\{ 0,\ldots,N\right\}$ for each $i\in\mathcal{N}$ and $m=\displaystyle \sum^{N}_{i=1} m_i$.\\
By assumption 
$m$ is odd.
Consider the following expected value
 \begin{eqnarray*}
\mathbb{E}\left[\displaystyle\prod^{N}_{i=1}\left(R_i - c_{i}\right)^{m_i}\right] =  \mathbb{E}\left[\displaystyle\prod^{N}_{i=1}\left(X_i \right)^{m_i}\right]=  \left(-1\right)^{m}\mathbb{E}\left[\displaystyle\prod^{N}_{i=1}\left(Y_i \right)^{m_i}\right]
 \end{eqnarray*}   
 by the equality in distribution \eqref{radsymeqdistr} $\mathbb{E}\left[\displaystyle\prod^{N}_{i=1}\left(Y_i \right)^{m_i}\right]=\mathbb{E}\left[\displaystyle\prod^{N}_{i=1}\left(X_i \right)^{m_i}\right]$.
 Then:
  \begin{eqnarray}\label{momeq}
\mathbb{E}\left[\displaystyle\prod^{N}_{i=1}\left(R_i - c_{i}\right)^{m_i}\right] = \left(-1\right)^{m}\mathbb{E}\left[\displaystyle\prod^{N}_{i=1}\left(R_i - c_{i}\right)^{m_i}\right]
 \end{eqnarray} 
Because $m$ is odd, the only possibility for equality \eqref{momeq} to hold is 
that the expected value is zero.\\
The case $m_i=m=1$ leads to i), ii) following from recognizing the odd moments in the zero expected values because of i).
ii) I define $w_i\in \mathbb{R}$ for each $i\in\mathcal{N}$ and $\mathbf{w}=\left(w_1,\ldots,w_N\right)$.\\ I denote the linear combination 
$R_p= \mathbf{w}^{\prime}\mathbf{R}$, with center  $c_p= \mathbf{w}^{\prime}\boldsymbol{c}$.\\ In addition, I call $X_p= \mathbf{w}^{\prime}\mathbf{X}$ and $Y_p= \mathbf{w}^{\prime}\mathbf{Y}$. 
\begin{eqnarray}
 F_{X_p}\left(x_p\right) = \mathbb{P}\left(X_p \leq x_p\right)=   \mathbb{P}\left(\mathbf{w}^{\prime}\mathbf{X} \leq x_p\right)= \mathbb{P}\left(-\mathbf{w}^{\prime}\mathbf{X} \geq -x_p\right)=\mathbb{P}\left(\mathbf{w}^{\prime}\mathbf{Y} \geq -x_p\right),
\end{eqnarray}
by \eqref{radsymeqdistr} $\mathbb{P}\left(\mathbf{w}^{\prime}\mathbf{Y} \geq -x_p\right)=\mathbb{P}\left(\mathbf{w}^{\prime}\mathbf{X} \geq -x_p\right)$, then
\begin{eqnarray}
F_{X_p}\left(x_p\right) =\mathbb{P}\left(\mathbf{w}^{\prime}\mathbf{Y} \geq -x_p\right) =\mathbb{P}\left(\mathbf{w}^{\prime}\mathbf{X} \geq -x_p\right) = \mathbb{P}\left(\mathbf{w}^{\prime}\mathbf{X} > -x_p\right) =\bar{F}_{X_p}\left(-x_p\right),
\end{eqnarray}
where I used the continuity of the multivariate CDF.
\end{proof}
\subsection{Proof of Proposition \ref{exced}}
\begin{proof}
Recalling that, under the central symmetry hypothesis, by i) in Proposition \ref{oddmom}, $\tilde{R}_i=\dfrac{X_i}{\sigma_i}$,  and $\tilde{R}_i{\buildrel d \over =} -\tilde{R}_i$, then
\begin{eqnarray*}
\mu^{+}_i\left(a\right) &=& \mathbb{E}\left(  \tilde{R}_i\right. \left\vert \tilde{R}_1> a ,\tilde{R}_2> a\right) = -  \mathbb{E}\left(  -\tilde{R}_i\right. \left\vert -\tilde{R}_1< -a ,-\tilde{R}_2< -a\right) \\
\sigma^{+}_i\left(a\right) &=& \sqrt{\mathbb{E}\left(  \left(\tilde{R}_i- \mu^{+}_i\left(a\right)\right)^2\right. \left\vert \tilde{R}_1> a ,\tilde{R}_2> a\right)}\\ &=& \sqrt{\mathbb{E}\left(  \left(-\tilde{R}_i+ \mu^{+}_i\left(a\right)\right)^2\right. \left\vert -\tilde{R}_1<- a ,-\tilde{R}_2< -a\right)}\\
\sigma^{+}_{1,2}\left(a\right)&=& \mathbb{E}\left( \left(\tilde{R}_1- \mu^{+}_1\left(a\right)\right) \left(\tilde{R}_2- \mu^{+}_2\left(a\right)\right) \right. \left\vert \tilde{R}_1> a ,\tilde{R}_2> a\right)
\\&=&  \mathbb{E}\left( \left(-\tilde{R}_1+ \mu^{+}_1\left(a\right)\right) \left(-\tilde{R}_2+ \mu^{+}_2\left(a\right)\right) \right. \left\vert -\tilde{R}_1< -a ,-\tilde{R}_2< -a\right)
\end{eqnarray*}
then under symmetry for $i=1,2$
\begin{eqnarray*}
-  \mathbb{E}\left(  -\tilde{R}_i\right. \left\vert -\tilde{R}_1< -a ,-\tilde{R}_2< -a\right)&=&-  \mathbb{E}\left(  \tilde{R}_i\right. \left\vert \tilde{R}_1< -a ,\tilde{R}_2< -a\right)
\end{eqnarray*}
i.e. $\mu^{+}_i\left(a\right)=-\mu^{-}_i\left(a\right)$, $i=1,2$\\
From the previous equation, under symmetry, also, the following result holds 
\begin{eqnarray*}
&&\sqrt{\mathbb{E}\left(  \left(-\tilde{R}_i+ \mu^{+}_i\left(a\right)\right)^2\right. \left\vert -\tilde{R}_1<- a ,-\tilde{R}_2< -a\right)}\\&=& \sqrt{\mathbb{E}\left(  \left(\tilde{R}_i- \mu^{-}_i\left(a\right)\right)^2\right. \left\vert \tilde{R}_1<- a ,\tilde{R}_2< -a\right)}.
\end{eqnarray*}
It follows then that $\sigma^{+}_i\left(a\right)=\sigma^{-}_i\left(a\right)$, $i=1,2$.\\
Analogously, central symmetry implies the following equation
\begin{eqnarray*}
&&\mathbb{E}\left( \left(-\tilde{R}_1+ \mu^{+}_1\left(a\right)\right) \left(-\tilde{R}_2+ \mu^{+}_2\left(a\right)\right) \right. \left\vert -\tilde{R}_1< -a ,\tilde{R}_2< -a\right)\\&=& \mathbb{E}\left( \left(\tilde{R}_1- \mu^{-}_1\left(a\right)\right) \left(\tilde{R}_2- \mu^{-}_2\left(a\right)\right) \right. \left\vert \tilde{R}_1< -a ,\tilde{R}_2< -a\right).
\end{eqnarray*}
The upper and lower covariances then satisfy $\sigma^{+}_{1,2}\left(a\right)=\sigma^{-}_{1,2}\left(a\right)$.\\
The result for exceedance correlations follows from their definitions.
\end{proof}
\subsection{Proof of Theorem \ref{margcop}}
\begin{proof}
\begin{eqnarray*}\label{H0}
F_{\mathbf{X}}\left(\mathbf{x}\right)&-&\bar{F}_{X}\left(-\mathbf{x}\right)=\\
C_{\mathbf{U}}\left(F_{X_1}\left(x_1\right),\ldots,F_{X_N}\left(x_N\right)\right) &-& \bar{C}_{\mathbf{U}}\left(\bar{F}_{X_1}\left(-x_1\right),\ldots,\bar{F}_{X_N}\left(-x_N\right)\right)=\\
C_{\mathbf{U}}\left(F_{X_1}\left(x_1\right),\ldots, F_{X_N}\left(x_N\right)\right) &-& \dfrac{1}{2}\bar{C}_{\mathbf{U}}\left(F_{X_1}\left(x_1\right),\ldots,F_{X_N}\left(x_N\right)\right) \\
+  \dfrac{1}{2}\bar{C}_{\mathbf{U}}\left(F_{X_1}\left(x_1\right),\ldots,F_{X_N}\left(x_N\right)\right)  &-&
\dfrac{1}{2}C_{\mathbf{U}}\left(\bar{F}_{X_1}\left(-x_1\right),\ldots,\bar{F}_{X_N}\left(-x_N\right)\right)\\+ \dfrac{1}{2} C_{\mathbf{U}}\left(\bar{F}_{X_1}\left(-x_1\right),\ldots,\bar{F}_{X_N}\left(-x_N\right)\right) &-&
\bar{C}_{\mathbf{U}}\left(\bar{F}_{X_1}\left(-x_1\right),\ldots,\bar{F}_{X_N}\left(-x_N\right)\right) 
\\= \Delta_{\mathbf{U}} \left(F_{X_1}\left(x_1\right),\ldots, F_{X_N}\left(x_N\right)\right) &+& C_{\mathbf{U}^{S}}\left(F_{X_1}\left(x_1\right),\ldots, F_{X_N}\left(x_N\right)\right) 
  \\ 
  - C_{\mathbf{U}^{S}}\left(\bar{F}_{X_1}\left(-x_1\right),\ldots,\bar{F}_{X_N}\left(-x_N\right)\right) &+& \Delta_{\mathbf{U}} \left(\bar{F}_{X_1}\left(-x_1\right),\ldots,\bar{F}_{X_N}\left(-x_N\right)\right)
\end{eqnarray*}

For the sufficiency of 1, let us call $u_i = F_{X_i}\left(X_i\right)$ and  $v_i = \bar{F}_{X_i}\left(-x_i\right)$ and   for $i \in \mathcal{N}$ for $i \in \mathcal{N}$.
Under marginal symmetry, I have $u_i = v_i$ and \eqref{radmarg}. For the necessity, since \eqref{radmarg} should be equal to zero every $\mathbf{x}\in \mathbb{R}^N$, I choose $\mathbf{x}$ such that all the components less the i-th component are $\infty$ obtaining
\begin{eqnarray*}
C_{\mathbf{U}^{S}}\left(1,\ldots, F_{X_i}\left(x_i\right),\ldots,1\right) &=&C_{\mathbf{U}^{S}}\left(1,\ldots,\bar{F}_{X_i}\left(-x_i\right),\ldots,1\right)  
\end{eqnarray*}
The previous equation is equivalent to :
\begin{eqnarray*}
\\ F_{X_i}\left(x_i\right)&=& \bar{F}_{X_i}\left(-x_i\right)
\end{eqnarray*}
The last line follows from $C_{\mathbf{U}^{S}}$ being a copula. The same could be shown for the other marginals with a proper choice of  $\mathbf{x}$.
For 2 as remarked above $C_{\mathbf{U}}$ is the CDF of $\mathbf{U}$ and $\bar{C}_{\mathbf{U}}$ is the CDF of  $\mathbf{1}_N- \mathbf{U}$ so  $\mathbf{U} {\buildrel d \over =} \mathbf{1}_N- \mathbf{U}$ is equivalent to $C_{\mathbf{U}}\left(\mathbf{u}\right) =\bar{C}_{\mathbf{U}}\left(\mathbf{u}\right)$ i.e. $\Delta_{\mathbf{U}}\left(\mathbf{u}\right)=0$ for every $\mathbf{u}\in\left[0,1\right]^N$.
Concerning 3 if \eqref{radmarg} and \eqref{radcop} are zero then $F_{\mathbf{}}\left(\mathbf{x}\right)=\bar{F}_{X}\left(-\mathbf{x}\right)$ and sufficency is estabilished. For necessity, by Corollary \ref{CS->MS}, marginal symmetry is a necessary condition for central symmetry. Then, marginal symmetry implies that \eqref{radmarg} is zero, and this implies that under central symmetry, \eqref{radcop} is also zero. By 2, copula central symmetry holds
\end{proof}

\subsection{Proof of Proposition \ref{empsurv}}
\begin{proof}
The result follows from the same line of reasoning of the proof of proposition 1 in \cite{quessy2016}. He considers the more general situation of the empirical copula process corresponding to the marginal  copula
\begin{eqnarray}
 C_{U_{\mathcal{I}},\mathcal{J}}= \mathbb{E}\left[\prod_{j \in \mathcal{I} \setminus \mathcal{J}} \mathbb{I}\left(U_{i_k}\leq u_{i_k}\right)\prod_{j \in  \mathcal{J}} \mathbb{I}\left(1-U_{i_k}\leq u_{i_k}\right)\right] 
\end{eqnarray}
but, restricting to i.i.d. data.
The survival copula case corresponds to $\mathcal{J}=\mathcal{I}=\mathcal{N}$ and
lemma 3 in \cite{quessy2016} implies the following relationships:
\begin{eqnarray}
\bar{C}_{\mathbf{U}}\left(\mathbf{u}\right) = \sum_{\mathcal{I}\subseteq \mathcal{N}} \left(-1\right)^{\left\vert \mathcal{I} \right\vert}  C_{\mathbf{U}_{\mathcal{I}}}\left(\mathbf{1}_{\left\vert \mathcal{I} \right\vert} -\mathbf{u}_{\mathcal{I}}\right)  
\\ \label{empsurrvrel}\bar{C}_{T}\left(\mathbf{u}\right) = \sum_{\mathcal{I}\subseteq \mathcal{N}} \left(-1\right)^{\left\vert \mathcal{I} \right\vert}  C_{T}\left(\mathbf{1}_{\left\vert \mathcal{I} \right\vert} -\mathbf{u}_{\mathcal{I}}\right) 
\end{eqnarray}
It follows that the survival copula process $\bar{\mathbb{C}}_T$ can be expressed as a linear combination of the empirical copula process $\mathbb{C}_T$ in equation \eqref{weaksurv}:
\begin{equation}
\bar{\mathbb{C}}_{T}\left(\mathbf{u}\right) = \sum_{\mathcal{I}\subseteq \mathcal{N}} \left(-1\right)^{\left\vert \mathcal{I} \right\vert}  \mathbb{C}_{T}\left(\mathbf{1}_{\left\vert \mathcal{I} \right\vert} -\mathbf{u}_{\mathcal{I}}\right)     
\end{equation}

Using  \eqref{weaksurv} from \cite{bucher2013consistent}, the continuous mapping theorem I obtain the limit
\begin{equation}\label{survlimit}
 \bar{\mathbb{C}}_{T}\leadsto \bar{\mathbb{C}} = \sum_{\mathcal{I}\subseteq \mathcal{N}} \left(-1\right)^{\left\vert \mathcal{I} \right\vert}  \mathbb{C}\left(\mathbf{1}_{\left\vert \mathcal{I} \right\vert} -\mathbf{u}_{\mathcal{I}}\right)        
\end{equation}

 Using the algebraic manipulations in the proof of proposition one in \cite{quessy2016}, the previous expression for $\bar{\mathbb{C}}$ becomes the one displayed in \eqref{weaksurv}.\\
Finally, I recognize in $\bar{\mathbb{B}}_{\mathbf{U}}$ the limit of the multivariate empirical process
\begin{equation}
\bar{\mathbb{B}}_{T,\mathbf{U}}=\dfrac{1}{\sqrt{T}}\sum^{T}_{t=1}\left\{ \mathbb{I}\left(\mathbf{1}_N-\mathbf{U}\leq \mathbf{u}\right)- \bar{C}_{\mathbf{U}}\left(\mathbf{u}\right)\right\}.    
\end{equation}

Under the assumption on the $\alpha$-mixing coefficient in the statement of the proposition and using theorem 7.3 in \cite{rio2017asymptotic}, it follows that  $\bar{\mathbb{B}}_{\mathbf{U}}$ is a d-dimensional Brownian sheet with covariance function in \eqref{survcov}.
\end{proof}

\subsection{Proof of Proposition \ref{statlim}}
\begin{proof}
Under the null, I rewrite  the antisymmetrization of the empirical copula as the antisymmetrization of the empirical copula process:
\begin{equation}
\sqrt{T}\Delta_{T}\left(\mathbf{u}\right)= \dfrac{\sqrt{T}}{2} \left( C_{T}\left(\mathbf{u}\right) - C_{\mathbf{U}}\left(\mathbf{u}\right) -\bar{C}_{T}\left(\mathbf{u}\right)+ \bar{C}_{\mathbf{U}}\left(\mathbf{u}\right)\right) = \dfrac{1}{2} \left( \mathbb{C}_{T}- \bar{\mathbb{C}}_{T} \right)  
\end{equation}
The result follows from the weak convergence of the Empirical Copula \citep{bucher2013}  and Survival Copula processes (Proposition \ref{empsurv})
, the application of the continuous mapping theorem to the map  $\xi: D\left([0,1]^d\right)\times D\left([0,1]^d\right)\mapsto \mathbb{R}$
\[\xi\left(\mathbb{A},\mathbb{B}\right)= \displaystyle\int_{\left(0,1\right]^d}\left(\mathbb{A}-\mathbb{B}\right)^2 dC\] and the $T$-consistency of the empirical copula as an estimator of the copula.
\end{proof}

\subsection{Proof of Proposition \ref{tie-breaksurv}}

\begin{proof}
By adapting the proof of lemma 3 in \cite{quessy2016} I derive the following result  
\begin{equation}
\bar{C}^{*,\eta}_{T}\left(\mathbf{u}\right) = \sum_{\mathcal{I}\subseteq \mathcal{N}} \left(-1\right)^{\left\vert \mathcal{I} \right\vert}  \bar{C}^{*,\eta}_{T}\left(\mathbf{1}_{\left\vert \mathcal{I} \right\vert} -\mathbf{u}_{\mathcal{I}}\right).  
\end{equation}

Then, using \eqref{empsurrvrel} the Tie-Break bootstrap survival empirical copula process 
can be expressed as a linear combination of the Tie-Break bootstrap empirical copula process $\mathbb{C}_T$ in equation \eqref{tiebreakprocess}:
\begin{equation}
\bar{\mathbb{C}}^{*,\eta}_{T}\left(\mathbf{u}\right) = \sum_{\mathcal{I}\subseteq \mathcal{N}} \left(-1\right)^{\left\vert \mathcal{I} \right\vert}  \mathbb{C}^{*,\eta}_{T}\left(\mathbf{1}_{\left\vert \mathcal{I} \right\vert} -\mathbf{u}_{\mathcal{I}}\right). \end{equation}

Finally, using  the continuous mapping theorem, \eqref{tiebreakprocess} and \eqref{survlimit} I obtain the statement of the proposition 
\begin{equation}
\bar{\mathbb{C}}^{*,\eta}_{T}\left(\mathbf{u}\right)\overset{\mathbb{P}}{\underset{ \mathbf{X}}{\,\leadsto}\,}\sum_{\mathcal{I}\subseteq \mathcal{N}} \left(-1\right)^{\left\vert \mathcal{I} \right\vert}  \mathbb{C}\left(\mathbf{1}_{\left\vert \mathcal{I} \right\vert} -\mathbf{u}_{\mathcal{I}}\right) = \bar{\mathbb{C}}\left(\mathbf{u}\right)  .  
\end{equation}
\end{proof}







\bibliographystyle{agsm}

\bibliography{Bibliography-MM-MC}

@article{1959fonctions,
 AUTHOR = {Sklar, M.},
     TITLE = {Fonctions de r\'{e}partition \`a {$n$} dimensions et leurs marges},
   JOURNAL = {Publ. Inst. Statist. Univ. Paris},
  FJOURNAL = {Publications de l'Institut de Statistique de l'Universit\'{e} de
              Paris},
    VOLUME = {8},
      YEAR = {1959},
     PAGES = {229--231},
}

@article{FDR,
AUTHOR = {Benjamini, Yoav and Hochberg, Yosef},
     TITLE = {Controlling the false discovery rate: a practical and powerful
              approach to multiple testing},
   JOURNAL = {J. Roy. Statist. Soc. Ser. B},
  FJOURNAL = {Journal of the Royal Statistical Society. Series B.
              Methodological},
    VOLUME = {57},
      YEAR = {1995},
    NUMBER = {1},
     PAGES = {289--300},
      ISSN = {0035-9246},
}

@article{nelsen1993some,
  AUTHOR = {Nelsen, Roger B.},
     TITLE = {Some concepts of bivariate symmetry},
   JOURNAL = {J. Nonparametr. Statist.},
  FJOURNAL = {Journal of Nonparametric Statistics},
    VOLUME = {3},
      YEAR = {1993},
    NUMBER = {1},
     PAGES = {95--101}
}

@book{nelsen2007introduction,
  AUTHOR = {Nelsen, Roger B.},
     TITLE = {An introduction to copulas},
    SERIES = {Springer Series in Statistics},
   EDITION = {Second},
 PUBLISHER = {Springer, New York},
      YEAR = {2006},
}

@article{genest2013tests,
  AUTHOR = {Genest, Christian and Ne\v{s}lehov\'{a}, Johanna G.},
     TITLE = {On tests of radial symmetry for bivariate copulas},
   JOURNAL = {Statist. Papers},
  FJOURNAL = {Statistical Papers},
    VOLUME = {55},
      YEAR = {2014},
    NUMBER = {4},
     PAGES = {1107--1119}
}

@article{bouzebda2012test,
  title = {Test of symmetry based on copula function},
journal = {Journal of Statistical Planning and Inference},
volume = {142},
number = {5},
pages = {1262-1271},
year = {2012},
author = {Salim Bouzebda and Mohamed Cherfi},
}

@article{longin2001extreme,
  title={Extreme correlation of international equity markets},
  author={Longin, Francois and Solnik, Bruno},
  journal={The Journal of Finance},
  volume={56},
  number={2},
  pages={649--676},
  year={2001},
  publisher={Wiley Online Library}
}

@article{ang2002asymmetric,
  title={Asymmetric correlations of equity portfolios},
  author={Ang, Andrew and Chen, Joseph},
  journal={Journal of Financial Economics},
  volume={63},
  number={3},
  pages={443--494},
  year={2002},
  publisher={Elsevier}
}

@article{hong2007asymmetries,
  title={Asymmetries in stock returns: Statistical tests and economic evaluation},
  author={Hong, Yongmiao and Tu, Jun and Zhou, Guofu},
  journal={Review of Financial Studies},
  volume={20},
  number={5},
  pages={1547--1581},
  year={2007},
  publisher={Soc Financial Studies}
}

@book{rio1999theorie,
   AUTHOR = {Rio, Emmanuel},
     TITLE = {Th\'{e}orie asymptotique des processus al\'{e}atoires faiblement
              d\'{e}pendants},
    SERIES = {Math\'{e}matiques \& Applications (Berlin) [Mathematics \&
              Applications]},
    VOLUME = {31},
 PUBLISHER = {Springer-Verlag, Berlin},
      YEAR = {2000},
     PAGES = {x+169}
}

@book{jondeau2007financial,
 AUTHOR = {Jondeau, Eric and Poon, Ser-Huang and Rockinger, Michael},
     TITLE = {Financial modeling under non-{G}aussian distributions},
    SERIES = {Springer Finance},
 PUBLISHER = {Springer-Verlag London, Ltd., London},
      YEAR = {2007},
     PAGES = {xviii+541}
}

@article{bucher2013consistent,
 AUTHOR = {B\"{u}cher, Axel and Ruppert, Martin},
     TITLE = {Consistent testing for a constant copula under strong mixing
              based on the tapered block multiplier technique},
   JOURNAL = {J. Multivariate Anal.},
  FJOURNAL = {Journal of Multivariate Analysis},
    VOLUME = {116},
      YEAR = {2013},
     PAGES = {208--229}
}

@book{van1996weak,
  AUTHOR = {van der Vaart, Aad W. and Wellner, Jon A.},
     TITLE = {Weak convergence and empirical processes},
    SERIES = {Springer Series in Statistics},
      NOTE = {With applications to statistics},
 PUBLISHER = {Springer-Verlag, New York},
      YEAR = {1996},
     PAGES = {xvi+508}
}

@article{dehgani2013measures,
  AUTHOR = {Dehgani, Azam and Dolati, Ali and \'{U}beda-Flores, Manuel},
     TITLE = {Measures of radial asymmetry for bivariate random vectors},
   JOURNAL = {Statist. Papers},
  FJOURNAL = {Statistical Papers},
    VOLUME = {54},
      YEAR = {2013},
    NUMBER = {2},
     PAGES = {271--286}
}

@article{Krupskii2016,
  AUTHOR = {Krupskii, Pavel},
     TITLE = {Copula-based measures of reflection and permutation asymmetry
              and statistical tests},
   JOURNAL = {Statist. Papers},
  FJOURNAL = {Statistical Papers},
    VOLUME = {58},
      YEAR = {2017},
    NUMBER = {4},
     PAGES = {1165--1187}
}

@book{joe2014dependence,
  title={Dependence Modeling with Copulas},
  author={Joe, Harry},
  series={Chapman \& Hall/CRC Monographs on Statistics \& Applied Probability},
  year={2014},
  publisher={Taylor \& Francis}
}

@article{quessy2016,
AUTHOR = {Quessy, Jean-Fran\c{c}ois},
     TITLE = {A general framework for testing homogeneity hypotheses about
              copulas},
   JOURNAL = {Electron. J. Stat.},
  FJOURNAL = {Electronic Journal of Statistics},
    VOLUME = {10},
      YEAR = {2016},
    NUMBER = {1},
     PAGES = {1064--1097}
}

@article{billio2021,
author = {Monica Billio and Lorenzo Frattarolo and Dominique Gu\'egan},
title = {Multivariate radial symmetry of copula functions: finite sample comparison in the i.i.d case},
journal = {Dependence Modeling},
number = {1},
volume = {9},
year = {2021},
pages = {43--61}
}

@article{beare2020, title={Randomization test of copula symmetry}, volume={36},  number={6}, journal={Econometric Theory}, publisher={Cambridge University Press}, author={Beare, Brendan K. and Seo, Juwon}, year={2020}, pages={1025–1063}}

@article{bucher2013,
title = {Empirical and sequential empirical copula processes under serial dependence},
journal = {Journal of Multivariate Analysis},
volume = {119},
pages = {61-70},
year = {2013},
author = {Axel Bucher and Stanislav Volgushev},
}

@book{durante2015principles,
  title={Principles of Copula Theory},
  author={Durante, F. and Sempi, C.},
  year={2015},
  publisher={CRC Press}
}

@article{Seo2024,
author = {Juwon Seo},
title = {Tie-Break Bootstrap for Nonparametric Rank Statistics},
journal = {Journal of Business \& Economic Statistics},
volume = {42},
number = {2},
pages = {615--627},
year = {2024},
publisher = {Taylor \& Francis}

}

@book{rio2017asymptotic,
  title={Asymptotic theory of weakly dependent random processes},
  author={Rio, Emmanuel},
  volume={80},
  year={2017},
  publisher={Springer}
}

@article{ohpatton,
title = {Dynamic factor copula models with estimated cluster assignments},
journal = {Journal of Econometrics},
volume = {237},
number = {2, Part C},
pages = {105374},
year = {2023},
author = {Dong Hwan Oh and Andrew J. Patton},
}

@article{Hong2006,
    author = {Hong, Yongmiao and Tu, Jun and Zhou, Guofu},
    title = "{Asymmetries in Stock Returns: Statistical Tests and Economic Evaluation}",
    journal = {The Review of Financial Studies},
    volume = {20},
    number = {5},
    pages = {1547-1581},
    year = {2006},
    month = {09},
}

@article{Albuquerque2012,
    author = {Albuquerque, Rui},
    title = "{Skewness in Stock Returns: Reconciling the Evidence on Firm Versus Aggregate Returns}",
    journal = {The Review of Financial Studies},
    volume = {25},
    number = {5},
    pages = {1630-1673},
    year = {2012},
    month = {01},   
}

@article{Bucher2014,
title = {Detecting changes in cross-sectional dependence in multivariate time series},
journal = {Journal of Multivariate Analysis},
volume = {132},
pages = {111-128},
year = {2014},
author = {Axel Bucher and Ivan Kojadinovic and Tom Rohmer and Johan Segers},
}

@article{joe2010,
title = {Tail dependence functions and vine copulas},
journal = {Journal of Multivariate Analysis},
volume = {101},
number = {1},
pages = {252-270},
year = {2010},
author = {Harry Joe and Haijun Li and Aristidis K. Nikoloulopoulos},
}

@article{Li2024,
title = {Multivariate directional tail-weighted dependence measures},
journal = {Journal of Multivariate Analysis},
volume = {203},
pages = {105319},
year = {2024},
author = {Xiaoting Li and Harry Joe}
}

@article{Hua2011,
title = {Tail order and intermediate tail dependence of multivariate copulas},
journal = {Journal of Multivariate Analysis},
volume = {102},
number = {10},
pages = {1454-1471},
year = {2011},
author = {Lei Hua and Harry Joe},
}

@article{Bormann2020,
author = {Carsten Bormann and Melanie Schienle},
title = {Detecting Structural Differences in Tail Dependence of Financial Time Series},
journal = {Journal of Business \& Economic Statistics},
volume = {38},
number = {2},
pages = {380--392},
year = {2020},
publisher = {ASA Website},
}

@article{Caballe1996,
title = {Mixed Risk Aversion},
journal = {Journal of Economic Theory},
volume = {71},
number = {2},
pages = {485-513},
year = {1996},
author = {Jordi Caball\'e and Alexey Pomansky},
}

@article{tsetlin2009,
  title={Multiattribute utility satisfying a preference for combining good with bad},
  author={Tsetlin, Ilia and Winkler, Robert L},
  journal={Management Science},
  volume={55},
  number={12},
  pages={1942--1952},
  year={2009},
  publisher={INFORMS}
}

@article{chen2015,
    author = {Chen, Yi-Ting},
    title = "{Exceedance Correlation Tests for Financial Returns}",
    journal = {Journal of Financial Econometrics},
    volume = {14},
    number = {3},
    pages = {581-616},
    year = {2015},
    month = {07},
}

@article{aki1993nonparametric,
  title={On nonparametric tests for symmetry in R m},
  author={Aki, Sigeo},
  journal={Annals of the Institute of Statistical Mathematics},
  volume={45},
  pages={787--800},
  year={1993},
  publisher={Springer}
}

@Article{billio2022,
AUTHOR = {Billio, Monica and Frattarolo, Lorenzo and Guégan, Dominique},
TITLE = {High-Dimensional Radial Symmetry of Copula Functions: Multiplier Bootstrap vs. Randomization},
JOURNAL = {Symmetry},
VOLUME = {14},
YEAR = {2022},
NUMBER = {1},
ARTICLE-NUMBER = {97},
}
\end{document}